\documentclass[%
 reprint,
superscriptaddress,
 amsmath,amssymb,
 aps,
 pra,
]{revtex4-2}
\usepackage{graphicx}
\usepackage{dcolumn}
\usepackage{bm}
\usepackage{hyperref}
\usepackage[mathlines]{lineno}
\usepackage{subcaption}
\usepackage{xcolor}
\usepackage{cleveref}
\usepackage{ragged2e}

\newcommand{\X}{\mathbb{B}}
\newcommand{\A}{A}

\begin{document}

\preprint{}

\title{Multipartite entanglement of random states of qubits}

\author{Giorgia Trotta}
\affiliation{Dipartimento di Fisica, Universit\`{a} di Napoli Federico II}
\affiliation{Dipartimento di Fisica, Università di Bari, 70126 Bari, Italy}
\affiliation{INFN, Sezione di Bari, 70125 Bari, Italy}

\author{Paolo Scarafile}
\affiliation{Dipartimento di Fisica, Universit\`{a} di Napoli Federico II}
\affiliation{Dipartimento di Fisica, Università di Bari, 70126 Bari, Italy}
\affiliation{INFN, Sezione di Bari, 70125 Bari, Italy}

\author{Paolo Facchi}
\affiliation{Dipartimento di Fisica, Università di Bari, 70126 Bari, Italy}
\affiliation{INFN, Sezione di Bari, 70125 Bari, Italy}

\author{Giuseppe Magnifico}
\affiliation{Dipartimento di Fisica, Università di Bari, 70126 Bari, Italy}
\affiliation{INFN, Sezione di Bari, 70125 Bari, Italy}

\author{Angelo Mariano}
\affiliation{Energy Technologies and Renewable Sources Department, ICT Division, ENEA (National Agency for New Technologies, Energy and Sustainable Economic Development), 70125 Bari, Italy}

\author{Giorgio Parisi}
\affiliation{Dipartimento di Fisica, Università degli Studi di Roma La Sapienza}
\affiliation{Istituto Nazionale di Fisica Nucleare, Sezione di Roma I}
\affiliation{Institute of Nanotechnology (NANOTEC) - CNR, Rome unit}

\author{Saverio Pascazio}
\affiliation{Dipartimento di Fisica, Università di Bari, 70126 Bari, Italy}
\affiliation{INFN, Sezione di Bari, 70125 Bari, Italy}

\author{ Karol Życzkowski}
\affiliation{Institute of Theoretical Physics, Faculty of Physics, Astronomy and Applied Computer Science, Jagiellonian University,
Kraków, Poland}
\affiliation{Center for Theoretical Physics (CFT), Polish Academy of Sciences,
Warszawa, Poland}

\begin{abstract}
We investigate multipartite entanglement via the statistical properties of pure quantum states of $n$-qubits. By analyzing the distribution of purity among balanced bipartitions, we compare Haar-typical states, uniformly distributed on the unit sphere of states, with Hadamard states, being characterized by equal weights in the computational basis. 
We analyze different ensembles of Hadamard states characterized by their phase distributions. Through analytical and numerical calculations, we show that Hadamard states exhibit, on average, a higher degree of entanglement than Haar-typical states. In addition, we show that a particular class of Hadamard states, characterized by real coefficients with alternating signs, known as hypergraph states, appears especially relevant in the search for maximally multipartite entangled states, both for their structural simplicity and the increased likelihood of sampling highly entangled states. These results identify Hadamard states as a tractable yet promising class for exploring multipartite entanglement structures and advancing the characterization of maximally multipartite entangled quantum states.

\end{abstract}

\maketitle
\section{\label{sec:level1}Introduction}
Entanglement is a fundamental feature of quantum mechanics and a central resource in quantum information, enabling relevant applications such as quantum teleportation~\cite{Bennett93,Bouwmeester97,Boschi98}, quantum computation~\cite{Nielsen00}, quantum communication~\cite{Horodecki01} and quantum cryptography~\cite{Bennett84, Ekert91, Ekert96, Fuchs97}. While the bipartite entanglement of pure states is unambiguously quantified in terms of the entropy
of either single-party reduced state~\cite{Wootters96,Wootters98,Wootters01}, the quantification of multipartite entanglement remains more intricate and less unified~\cite{Bruss02,Amico08,Horodecki09}. This has led to multiple approaches being developed, each focusing on different aspects of the problem, for instance the invariance under permutation of the parties~\cite{Coffman00, Wong01}, the scalability of the measure with the system size~\cite{Meyer02}, the structural analogy with the bipartite von Neumann entropy~\cite{Bruss02} or the mutual exclusivity of single-partite and bipartite properties of subsystems~\cite{Jakob07}. These different characterizations of multipartite entanglement often yield distinct and, in some cases, incompatible classifications.

Various studies have shown that multipartite entanglement exhibits frustration, i.e., the impossibility of simultaneously maximizing entanglement across all bipartitions ~\cite{Rains99, Higuchi00,Scott04,Huber17,Facchi_frustration}, making statistical methods from complex systems particularly suitable for its characterization. 

In this context, analyzing the statistical distribution of entanglement measures, such as the purity over all balanced bipartitions and related quantities like the average linear entropy~\cite{Bruss02,Scott04,Facchi08}, has emerged as a powerful tool for characterizing multipartite entanglement~\cite{Facchi06}. This approach has been effectively applied to typical pure states sampled according to the Haar measure in qubit systems~\cite{Giraud07,Facchi10,Facchi09.qubit,Facchi_classical}.

A particularly significant class of states is that of maximally multipartite entangled states (MMES), which are defined as those that maximize entanglement across all balanced bipartitions~\cite{Facchi08}. The presence of quantum frustration makes the identification of these maximally entangled states particularly demanding and significant. 
In addition, the number of free parameters in quantum states and the number of bipartitions of the total system both grow exponentially with system size,  making the investigation of multipartite entanglement and the search for MMES analytically and computationally very challenging~\cite{Facchi_frustration,Facchi10}.

In this framework, \emph{Hadamard states},   so named as they corresponds to a single row (or column) of a complex Hadamard matrix of the Butson class~\cite{Butson62,Zyc06}, and characterized by equal weights in the computational basis, have been identified as compatible with MMES requirements, providing a structured class for theoretical and computational analysis~\cite{Facchi09.qubit}. Their intriguing properties, in terms of frustration and symmetries, make them promising candidates for investigating multipartite entanglement in qubit systems.

In this work, we characterize the entanglement properties of Hadamard states of $n$ qubits using a statistical approach. We analyze different ensembles of Hadamard states, having equal weights in the computational basis and phases independently and uniformly distributed over the $q$-th roots of  unity, with $q=2,3,\dots$: \emph{hypergraph states} ($q=2$), introduced in Ref.~\cite{Macchiavello13}, whose phases are
the square roots of unity, either $1$ or $-1$ and which correspond to a row (or column) of a real Hadamard matrix of the Butson class; the \emph{Hadamard-Butson $P_q$-states} ($q=3,4,\dots$), whose possible phases are the vertices  of a regular $q$-polygon, namely, $e^{i\frac{2\pi r}{q}}$, with $r=0,\dots q-1$;
and \emph{Hadamard-typical states} ($q=\infty$),
whose phases are uniformly distributed over the unit circle. Notice that Hadamard-typical states are locally maximally entanglable states (LMEs), as showed in Ref.~\cite{Kraus08}.

Through analytical calculations and numerical simulations, we demonstrate that Hadamard ensembles exhibit, on average, a higher degree of entanglement than the Haar ensemble. Moreover, we show that the class of hypergraph states exhibits distinctive statistical properties that make them particularly promising for identifying MMES.
In addition, hypergraph states occur naturally in the analysis of quantum algorithms
such as the Deutsch-Jozsa algorithm and the Grover algorithm~\cite{Macchiavello13,Macchiavello14} and, can be complex in the sense of Kolmogorov complexity, so they could, for instance, be used for
quantum fingerprinting protocols~\cite{Kraus07}. Nonetheless, identifying MMES, even in the class of hypergraph states remains challenging, particularly as the system size increases.

The Article is structured as follows. In Sec.~\ref{sec2:preliminaries}, we introduce the necessary background and notation, review key results on Haar-typical states, and define the classes of Hadamard states under investigation. Section \ref{sec3:fixed bipartition} is devoted to the evaluation of the first two cumulants of the purity distribution associated to a fixed bipartition, while Sec.~\ref{sec4:MEcumulants} extends the analysis to the purity averaged over all balanced bipartitions. Throughout, particular emphasis is placed on comparing the statistical entanglement features of Hadamard and Haar ensembles.
Lastly, in Sec.~\ref{sec:asymp}, we analyze the large-system limit of the purity distributions for Hadamard states and compare it with that of Haar-typical states.

\section{\label{sec2:preliminaries}Preliminaries and Definitions}

We consider a system \(S = \{1, \ldots, n\}\) of \(n\) qubits, described by a Hilbert space of dimension \(N = 2^n\),
\begin{equation}
\mathcal{H}_\mathcal{S} = \bigotimes_{i \in \mathcal{S}} \mathbb{C}^2_i \simeq \mathbb{C}^{N}
\end{equation}
A pure state of the system can be expressed in the computational basis as
\begin{equation}
|\psi\rangle = \sum_{k \in \X^n} z_k |k\rangle,
\label{eq:purestate}
\end{equation}
where \(\X = \{0,1\}\), \(k = (k_1, \ldots, k_n)\) with \(k_i \in \X\), \mbox{$|k\rangle = \bigotimes_{i \in \mathcal{S}} |k_i\rangle_i$}, \(|k_i\rangle_i \in \mathbb{C}^2_i\), and the normalization condition \(\sum_k |z_k|^2 = 1\) holds.

A bipartition of the system is defined by a pair \((\A, \bar{\A})\), where \(\A \subset {S}\), \(\bar{\A} = {S} \setminus \A\), and \(1 \leq n_{\A} \leq n_{\bar{\A}}\), with \mbox{\(n_{\A} = |\A|\)}. A bipartition is said to be \emph{balanced} if \mbox{\(n_{\A} = \lfloor n/2 \rfloor\)}.

Given a bipartition \((\A, \bar{\A})\) and a pure state $|\psi\rangle$, the purity of subsystem \(\A\) is defined as
\begin{equation}
\pi_{\A} = \mathrm{Tr}(\rho_{\A}^2), \qquad \rho_A = \mathrm{Tr}_{\bar{{\A}}}(|\psi\rangle\langle\psi|).
\end{equation}
The purity quantifies the degree of entanglement between the two subsystems: lower values of \(\pi_{\A}\) indicate stronger entanglement. For a subsystem of $n_\A$ qubits,
the purity is bounded as
\begin{equation}\label{eq:purity_bounds}
\frac{1}{2^{n_{\A}}} \le \pi_{\A} \le 1.
\end{equation}

The purity of a fixed bipartition $(\A,\bar{\A})$ has the following expression~\cite{Facchi06}:
\begin{equation}
    \pi_{\A}(\bm{z}) = \sum_{k\in \X^n} \sum_{l\in \X^{\A}} \sum_{m\in \X^{\bar{\A}}}z_k z^*_{k\oplus l} z_{k\oplus l \oplus m}z^*_{k\oplus m},
    \label{def_piA}
\end{equation}
where $\oplus$ is the bitwise XOR operation.
This quantity describes the entanglement of the single bipartition $(\A,\bar{\A})$, and fails in capturing the entanglement properties of the system as a whole.

A quantification of global multipartite entanglement for pure states is provided by the \emph{potential of multipartite entanglement}, defined as the average purity over all balanced bipartitions~\cite{Facchi08},
\begin{equation}
    \pi_{\mathrm{ME}} = \binom{n}{\lfloor \frac{n}{2} \rfloor}^{-1} \sum_{\substack{|{\A}|= \lfloor \frac{n}{2} \rfloor}} \pi_{\A}.
    \label{eq: multpotentialdef}
\end{equation}
This quantity can be expressed as follows in terms of the state's Fourier coefficients $\bm{z}=(z_1,\dots,z_N)$~\cite{Facchi08,Facchi09.qubit}:
\begin{equation}\label{def:purityME}
    \pi_{\mathrm{ME}}(\bm{z})=\sum_{k,l,m\in \X^n}g(l,m)z_k z^*_{k\oplus l} z_{k\oplus l \oplus m} z^*_{k\oplus m} ,
\end{equation}
where
\begin{subequations}
    \begin{equation}
        g(l,m) = \delta_{l \land m,0} \, \hat{g}(|l|, |m|), \label{gdef}
    \end{equation}
    \begin{equation}
        \hat{g}(s,t) = \frac{1}{2} \binom{n}{\lfloor \frac{n}{2} \rfloor}^{-1} \left[ \binom{n - s - t}{\lfloor \frac{n}{2} \rfloor - s} + \binom{n - s - t}{\lfloor \frac{n}{2} \rfloor - t} \right]. \label{gbardef}
    \end{equation}
\end{subequations}
Here, \(|l| = \sum_{i \in \mathcal{S}} l_i\), while \(\land\) denotes the bitwise AND operation.

The average purity $\pi_{\mathrm{ME}}$ inherits the bounds~\eqref{eq:purity_bounds} of the purity $\pi_A$ of a fixed balanced bipartition, with $n_A=\lfloor n/2 \rfloor$, namely,
    \begin{equation}\label{eq:pME_bounds}
\frac{1}{2^{\lfloor \frac{n}{2} \rfloor}} \le \pi_{\mathrm{ME}} \le 1.
\end{equation}

Correspondingly, \emph{maximally multipartite entangled states} (MMES) are defined as those minimizing the potential of multipartite entanglement (\ref{eq: multpotentialdef}). 

In the case of a system of $n$ qubits the
minimal possible value of the potential of multipartite entanglement $\pi_{\mathrm{ME}}$ equals $2^{-\lfloor n/2 \rfloor}$. This  theoretical bound for the minimum is saturated if perfect MMES or \textit{Absolutely Maximally Entangled} (AME) states~\cite{Zyczkowsi15,Zyczkowsi18,Zyczkowsi21,Zyczkowsi22,Zyczkowski23} exist.
 For $n = 2$ AME states are Bell states up to local unitary transformations, while for $n = 3$ they include the GHZ states~\cite{Facchi08,Zyczkowsi18}. For $n = 5$ and $n = 6$, numerous examples of AME states have been identified~\cite{Facchi08}. Conversely, it has been proven that AME states do not exist for $n = 4$~\cite{Higuchi00}, $n = 7$~\cite{Rains99,Scott04,Huber17}, and $n \geq 8$~\cite{Rains99,Scott04}, due to  quantum frustration \cite{Facchi_frustration}. Frustration arises due to competing requirements among different bipartitions.

In the following we introduce Haar-typical and Hadamard states and review the main properties of the former.

\subsection{Statistical Properties of Haar-Typical States}
The unit sphere of states~\eqref{eq:purestate} of $n$ qubits in \(\mathcal{H}_\mathcal{S}\),
\begin{equation}
    \mathbb{S}^{2N-1} = \Bigl\{ \bm{z}=(z_1, \ldots, z_N) \in \mathbb{C}^N \,:\, \sum_k |z_k|^2 = 1 \Bigr\},
    \label{eq:sphere}
\end{equation}
 is invariant under the action of the unitary group \(\mathrm{U}(\mathcal{H}_\mathcal{S})\simeq \mathrm{U}(N)\), with $N=2^n$, and Haar-typical pure states are sampled uniformly on $\mathbb{S}^{2N}$, according to the Haar measure of $\mathrm{U}(N)$~\cite{Zyczkowski01}.

The properties of Haar-typical pure states have been extensively studied~\cite{Lubkin78, Lloyd88, Page93, Zyczkowski01, Scott03, Giraud07, Facchi06, Facchi07,Facchi_classical}, and such states can be efficiently generated through chaotic dynamics~\cite{Cenedese23}.
The distribution of the purity referred to a fixed  bipartition $(\A,\bar{\A})$  
has mean and variance given by~\cite{Lloyd88,Facchi06,Giraud07}
    \begin{equation}
        \mu_{\A}  = \langle \pi_{\A} \rangle = \frac{2^{n_\A} + 2^{n_{\bar{\A}}} }{2^n+1}, 
        \label{mutypA1}
    \end{equation}
    \begin{equation}
        \sigma^2_\A  = \langle (\pi_{\A} - \mu_{\A})^2 \rangle = \frac{2\left(2^{2 n_\A} - 1\right)\left(2^{2n_{\bar{\A}}} - 1\right)}{(2^n + 1)^2(2^n + 2)(2^n + 3)}, 
        \label{sigmatypA}
    \end{equation}
where 
$n_\A=|\A|$ is the number of qubits in subsystem $\A$,   $n_{\bar{\A}}=n-n_\A$, and
\(\langle \cdots \rangle\) denotes the ensemble average over pure states uniformly distributed on $\mathbb{S}^{2 N-1}$, 
\begin{equation}
   \bm{z}\sim \mathrm{Unif}(\mathbb{S}^{2 N-1}). 
   \label{eq:Haarensemble}
\end{equation}

The mean of $\pi_{\textrm{ME}}$ is 
\begin{equation} \label{typicalMeanME}
    \mu_{\mathrm{ME}} = \langle \pi_{\mathrm{ME}} \rangle = \mu_\A,
\end{equation}
with $(\A,\bar{\A})$ being any balanced bipartition, and thus it is given by~\eqref{mutypA1} for $n_\A=\lfloor n/2 \rfloor$,
\begin{align}
        \mu_{\mathrm{ME}}  &= \frac{2^{\lfloor \frac{n}{2} \rfloor} + 2^{\lceil \frac{n}{2} \rceil} }{2^n+1}.
        \label{mutypA}
    \end{align}
The variance, $\sigma^2_{\mathrm{ME}} = \langle \left( \pi_{\mathrm{ME}} - \mu_{\mathrm{ME}} \right)^2\rangle$, is given by~\cite{Facchi_classical}
\begin{align} \label{typicalSigmaME}
    \sigma^2_{\mathrm{ME}}
    =\frac{(2^n + 1) f_2(n) - 2\left(2^{\lfloor\frac{n}{2}\rfloor} + 2^{\lceil\frac{n}{2}\rceil}\right)^2}{(2^n + 1)^2 (2^n + 2)(2^n + 3)},
\end{align}
where
\begin{align} \label{def_f2}
    f_2(n) &= 4 \sum_{k,l\in\X^n} g(k,l)^2
    \nonumber\\
    &=\sum_{s,t=0}^{\lceil\frac{n}{2}\rceil} \binom{n}{s,t}^{\!\!\!-1} \!\left[ 
    \binom{\lfloor\frac{n}{2}\rfloor}{s} \binom{\lceil\frac{n}{2}\rceil}{t} 
    + \binom{\lfloor\frac{n}{2}\rfloor}{t} \binom{\lceil\frac{n}{2}\rceil}{s} \right]^2
\end{align}
with $\binom{n}{s,t}= \frac{n!}{s! t! (n-s-t)!}$ being the trinomial coefficient.

\subsection{Hadamard States}

For any  pure state expressed in the computational basis, one can define the population probability vector $(P_{S}(k))_{k\in\X^n} = (|z_k|^2)_{k\in\X^n}$,
whose components represent the probability of observing the computational basis state  $|k\rangle$.
The population probability vector of Haar-typical states~\eqref{eq:Haarensemble} is uniformly distributed on the $(N-1)$-dimensional probability simplex $\Delta_{N-1}$, namely $P_S \sim \mathrm{Uniform}(\Delta_{N-1})$~\cite{Bengtsson2017}.  

A necessary condition for a state to qualify as an AME state, and hence to saturate the theoretical minimum in~\eqref{eq:pME_bounds}, is that all the marginals over $n_\A\leq n/2$ variables of its population
probability vector, \(P_{{S}}(k)\), are completely
random~\cite{Facchi09.qubit}. 

A notable class of states satisfying this condition is the class of \emph{Hadamard states}, defined by the property
\begin{equation}
    |z_k|^2 = \frac{1}{2^n}, 
    \qquad \forall k \in \X^n,
    \label{eq:AMEnec}
\end{equation}
which fixes the population probability vector at the center of the simplex $\Delta_{N-1}$.
Thus, for Hadamard states, the Fourier coefficients take the form,
\begin{equation}
    z_k = \frac{s_k}{\sqrt{2^n}},
    \qquad s_k = e^{i\phi_k} \in \mathbb{S}, \qquad
    \forall k\in \X^n,
    \label{defUnif}
\end{equation}
with $\mathbb{S}=\{z\in\mathbb{C}\,:\, |z|=1\}$ being the unit circle of complex numbers of unit modulus.

A central objective in the study of multipartite entanglement is the characterization of MMES, particularly in the context of frustrated systems, where AME states do not exist. However, MMES states are highly atypical and occur with  vanishing probability under Haar measure~\cite{Facchi08}. In the case of \(n\) qubits, a generic pure state is described by \(N = 2^n\) complex amplitudes \(\bm{z}=(z_1, \ldots, z_N)\), that correspond to a point on the $(2N-1)$-dimensional sphere~\eqref{eq:sphere}. This results in a high-dimensional optimization problem when attempting to minimize the average purity.

To address this challenge, it is useful to investigate classes of quantum states characterized by a reduced number of parameters, while still satisfying the necessary condition for perfect MMES. It is known that Haar-typical states have Fourier coefficients with independently  and uniformly distributed phases  \(s_k=z_k/|z_k|\sim\mathrm{Unif}(\mathbb{S})\)~\cite{Facchi10,Bengtsson2017}. The approach adopted in this work retains this statistical feature 
while restricting the number of free parameters.

This motivates the study of the statistical properties of Hadamard states, particularly those whose phases \(s_k\) are in a finite number and symmetrically distributed on the unit circle $\mathbb{S}$. We refer to these as Hadamard-Butson $P_q$-states, for $q=2,3,\dots$, and they are
listed in Table~\ref{tab:table1}. They are classified according to the allowed phases $s_k\in P_q$, taken to be the vertices of the regular polygon 
\begin{equation}
    P_q = \left\{e^{i\frac{2\pi  r}{q}} \ : \ {r=0, \dots,q-1} \right\}
\end{equation}
of the $q$-th roots of  unity. In particular, the Butson $P_2$-states coincide with the hypergraph states, whose phases are restricted to $\pm 1$.

In the following analysis we will consider, as ensembles of random states, Hadamard-Buston $P_q$-states with phases independently and uniformly distributed over the polygon, 
\begin{equation}
    s_k\sim \mathrm{Unif}(P_q), \qquad k\in\X^n ,
    \label{eq:Butsonq}
\end{equation}
together with the limiting case $q=\infty$, corresponding  to Hadamard-typical states, whose  
phases are independently and uniformly distributed over the entire unit circle, 
\begin{equation}
    s_k\sim\mathrm{Unif}(\mathbb{S}), \qquad k\in\X^n.
    \label{eq:Hadamardtyp}
\end{equation}
We will compare their statistical properties with those of the Haar-typical states discussed in the previous section.

\begin{table}[t]
    \centering
    \begin{tabular}{c|c}
        \hline
        \rule{0pt}{3mm} Fourier coefficients $z_k = \frac{s_k}{\sqrt{2^n}}$ & Notation \\[1mm]
        \hline
        \rule{0pt}{4mm}
        \( s_k \in P_2 =\{+1, -1\}\) & Hypergraph states ($q=2$) \\[2mm]
        \(s_k \in P_q=\left\{ e^{i\frac{2\pi  r}{q}} \right\}_{r=0, \dots,q-1}\) & 
        Butson \(P_q\)-states ($q \geq 3$) \\[2mm]
        \(s_k \in \mathbb{S} = \left\{ e^{i \phi} \right\}_{\phi \in [0,2\pi]} \) & 
        Hadamard states ($q=\infty$) \\[2mm]
        \hline
    \end{tabular}
\caption{\justifying \label{tab:table1} Classification of Hadamard-Butson $P_q$-states ($q =2,3,\dots$) based on the possible values of the phases $s_k\in\mathbb{S}$
of their Fourier coefficients~\eqref{defUnif}
 in the computational basis.}
\end{table}

\section{\label{sec3:fixed bipartition}First two cumulants of the Purity at Fixed Bipartition}

\begin{figure*}[t]
    \centering
    \begin{subfigure}[h]{0.32\textwidth}
        \includegraphics[width=\textwidth]{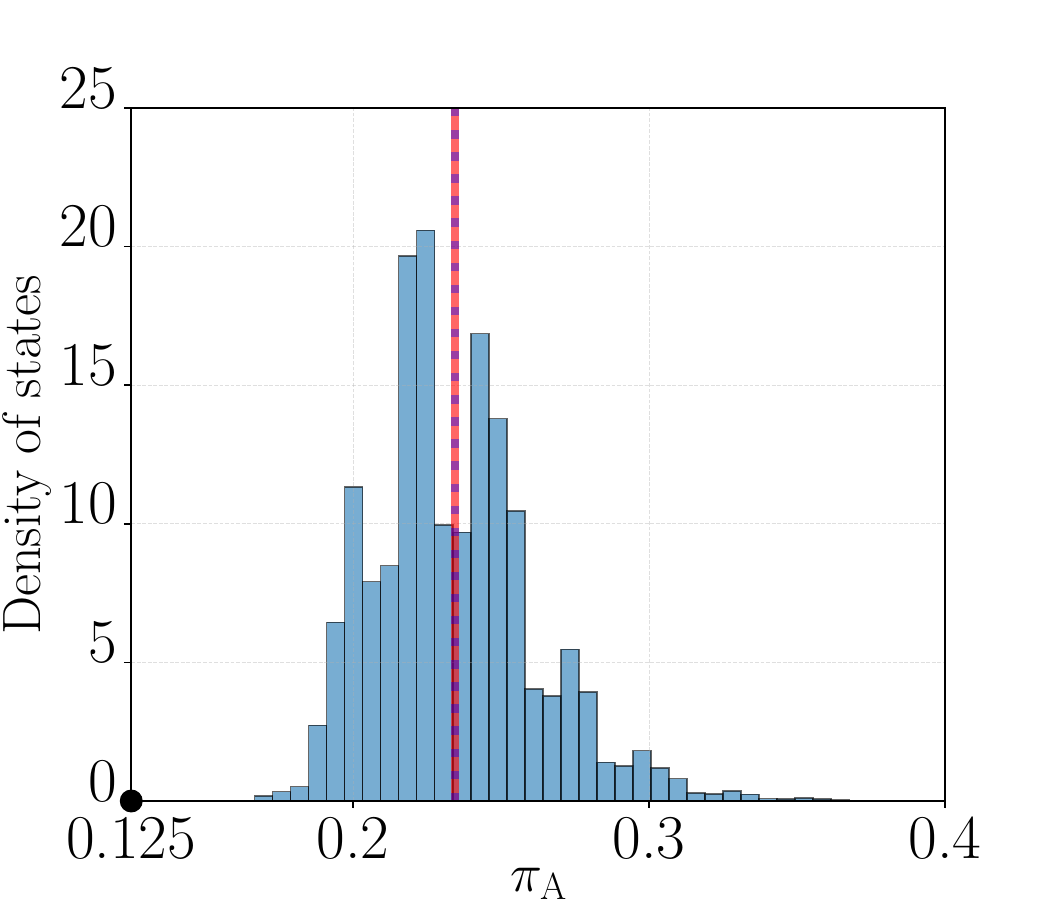}
        \caption{}
        \label{subfig:pma6}
    \end{subfigure}
    \begin{subfigure}[h]{0.32\textwidth}
        \includegraphics[width=\textwidth]{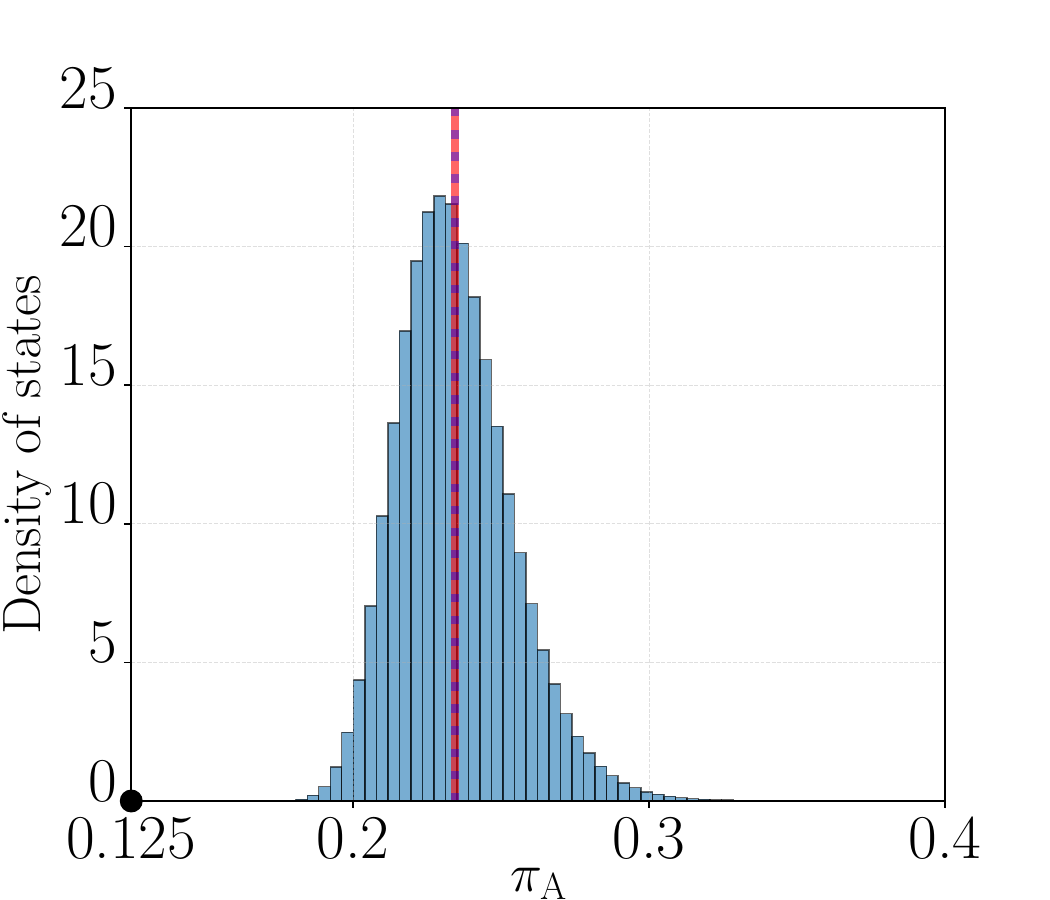} 
        \caption{}
        \label{subfig:P4a6}
    \end{subfigure}
    \begin{subfigure}[h]{0.32\textwidth}
        \includegraphics[width=\textwidth]{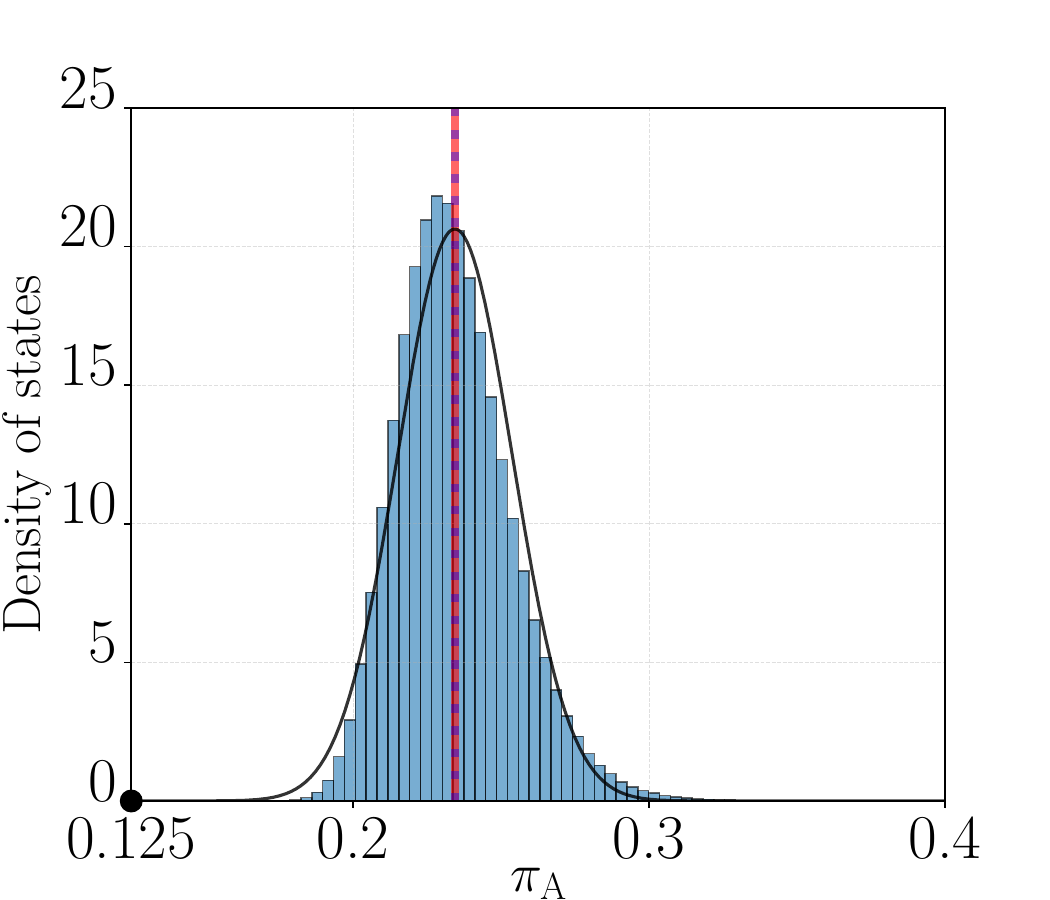} 
        \caption{}
        \label{subfig:randa6}
    \end{subfigure}
        \begin{subfigure}[h]{0.32\textwidth}
        \includegraphics[width=\textwidth]{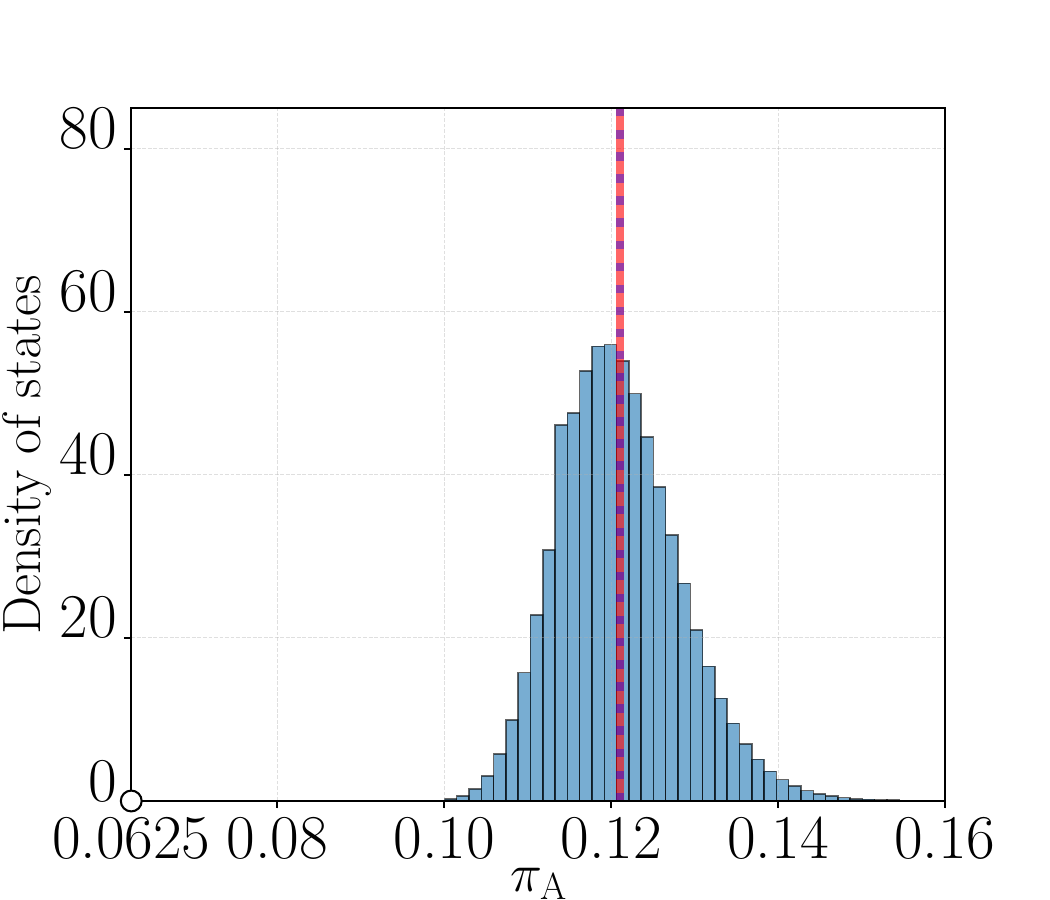}
        \caption{}
        \label{subfig:pma8}
    \end{subfigure}
    \begin{subfigure}[h]{0.32\textwidth}
        \includegraphics[width=\textwidth]{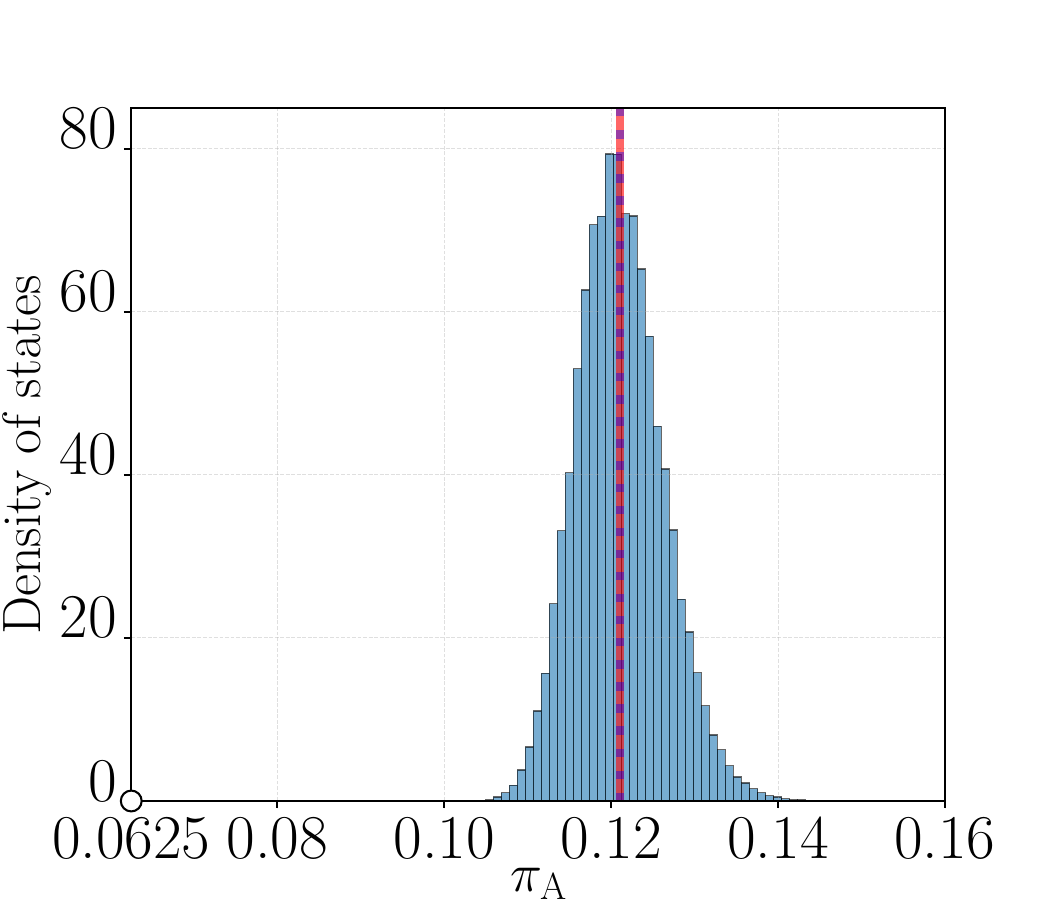} 
        \caption{}
        \label{subfig:P4a8}
    \end{subfigure}
    \begin{subfigure}[h]{0.32\textwidth}
        \includegraphics[width=\textwidth]{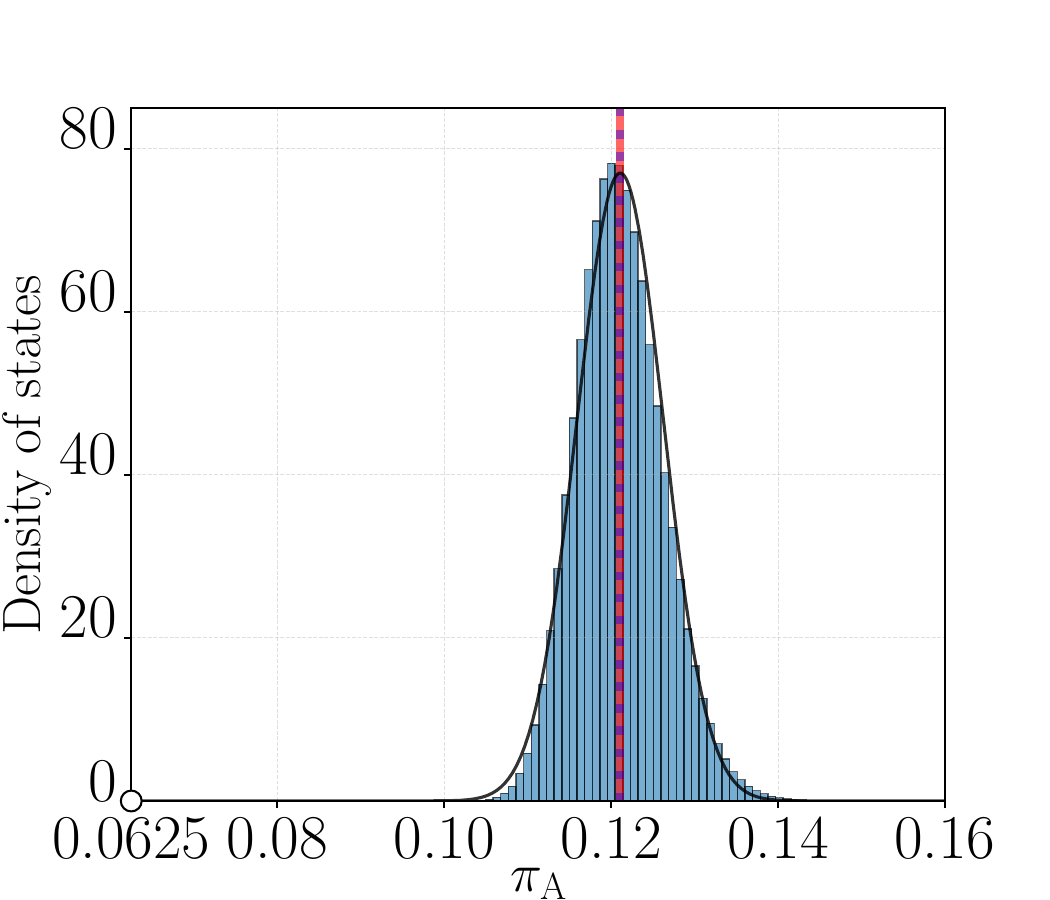} 
        \caption{}
        \label{subfig:randa8}
    \end{subfigure}
    \caption{\justifying Histograms of the purity distribution for the balanced bipartition $(\A,\bar{\A})$ with $\A=\{1,...,\lfloor n/2\rfloor\}$, based on \(2.3 \times 10^6\) samplings of systems of $n=6$ qubits (first row) and $n=8$ qubits (second row) for Hadamard-Butson $P_q$-states: (\subref{subfig:pma6},\subref{subfig:pma8}) hypergraph states ($q=2$); (\subref{subfig:P4a6},\subref{subfig:P4a8}) Butson $P_4$-states and (\subref{subfig:randa6},\subref{subfig:randa8}) Hadamard-typical states ($q=\infty$). In all the plots, the horizontal axis starts from the minimum value of $\pi_{\mathrm{A}}$. Red lines indicate the theoretical value of the mean purity~\eqref{meanA}, while the (almost coinciding) dotted blue lines correspond to the numerically computed one. In \subref{subfig:randa6},\subref{subfig:randa8} we have also plotted the Gaussians having mean and variance corresponding to the ones in Eqs.~\eqref{meanA} and~\eqref{varianceA}. We observe that the Gaussian approximation already provides an accurate description for $n=8$.
    Notice that for $n=6$ qubits, in the first line, there exist AME Hadamard states, with all purities of balanced bipartitions at the minimal value $1/8=0.125$, while for $n=8$ qubits, in the second line, there exist no AME state, with all $\pi_\A$ having the minimal value $1/16=0.0625$, as highlighted by the full and empty dots on the horizontal axis, respectively. 
}
    \label{fig:pmA}
\end{figure*}

In this section, we investigate the first two cumulants of the purity distribution associated with the classes of random Hadamard states listed in Table~\ref{tab:table1}.

As shown in Ref.~\cite{Facchi09.qubit}, the purity corresponding to a fixed bipartition \((\A, \bar{\A})\) of a Hadamard state can be expressed as

\begin{align}
\pi_{\A}(\bm{s}) &= \frac{2^{n_\A} + 2^{n_{\bar{\A}}} - 1}{2^{n}}  \nonumber \\
&\quad + \frac{1}{2^{2n}} \sum_{k \in \X^n} \sum_{l \in \X_*^{\A}} \sum_{m \in \X_*^{\bar{\A}}} 
s_k s_{k \oplus l}^* s_{k \oplus l \oplus m}  s_{k \oplus m}^*,
\label{defPIa}
\end{align}
where $\X^m_* =\X^m \setminus\{0\}$ denotes the set of all nonzero $m$-bit binary strings.
Importantly, since their indexes are all distinct, each term in the sum  of Eq.~\eqref{defPIa} consists of four independent random variables of zero mean~\eqref{eq:Butsonq}, and thus has zero average. Therefore, the mean of the purity distribution is given by the first term in~\eqref{def_piA}, i.e.\
\begin{equation}
\mu_{\A}=\langle \pi_{\A} \rangle = \frac{2^{n_{\A}} + 2^{n_{\bar{\A}}} - 1}{2^n}.
\label{meanA}
\end{equation}
This allows us to write the purity as a sum of its average and a fluctuation term,
\begin{equation}
\pi_{\A}(\bm{s}) = \mu_{\A} + \widetilde{\pi}_{\A}(\bm{s}),
\end{equation}
with \(\widetilde{\pi}_{\A}\) being given by the second term in~\eqref{defPIa}.

Consequently, the variance of the purity distribution is \(\sigma_A^2 = \langle \widetilde{\pi}_A^2 \rangle\) and is explicitly evaluated in Appendix~\ref{Appendix_variances}, yielding
\begin{equation}\label{varianceA}
    \sigma_{\A}^2 = 2 c_q \frac{(2^{n_{\A}} - 1)(2^{n_{\bar{\A}}} - 1)}{2^{3n}},
\end{equation}
where 
\begin{equation}
    c_q = (1+\delta_{q,2}),
\end{equation}
for all
Butson $P_q$-states~\eqref{eq:Butsonq} with $q\geq 2$, including Hadamard-typical states~\eqref{eq:Hadamardtyp} ($q=\infty$).

Comparing the mean purities of Hadamard~\eqref{meanA} and Haar ensembles~\eqref{mutypA1}, one finds that, for every bipartition~$(\A,\bar{\A})$,  
\begin{equation}
	(\mu_\A)_{\mathrm{Hadamard}} 
    <(\mu_{\A})_{\mathrm{Haar}}.
\end{equation}
Thus, Hadamard states exhibit, on average, a higher degree of entanglement than Haar states:
imposing the  condition~\eqref{eq:AMEnec} selects a subset of more entangled states.

Incidentally, observe that Eq.~\eqref{meanA} generalizes to the full class of hypergraph states the results previously obtained in~\cite{Hamma22} via a tensor-network approach. In particular, it shows that the mean bipartite purity for hypergraph states coincides with that of graph states. Furthermore, Eq.~\eqref{varianceA} reveals that the fluctuations of bipartite entanglement across the ensemble of random hypergraph states exhibit the same scaling behavior identified in Ref.~\cite{Hamma22} for a specific subclass of this ensemble composed by purely hypergraph states, while scaling significantly faster than the fluctuations associated exclusively with graph states.

We benchmark the analytical results through extensive numerical simulations performed as described in Appendix~\ref{Appendix_simulations}. The numerical data are shown in~\Cref{subfig:pma6,subfig:pma8} for hypergraph states, in~\Cref{subfig:P4a6,subfig:P4a8} for Butson $P_4$-states, and in~\Cref{subfig:randa6,subfig:randa8} for Hadamard-typical states. The results are in close agreement with the theoretical predictions given by Eqs.~\eqref{meanA} and \eqref{varianceA}.
Fig.~\ref{fig:pmA} also highlights the discreteness of the purity spectra associated with the Butson $P_q$-states. However, as the qubit number increases, the purity distribution approximates a continuous one.

\section{\label{sec4:MEcumulants}First two cumulants of the average purity}
We now extend the previous analysis to the distribution of the average purity over all balanced bipartitions~\eqref{eq: multpotentialdef}. Using the approach of Ref.~\cite{Facchi09.qubit}, the average purity for Hadamard $P_q$-states can be expressed as
\begin{align} \label{ExpressionPiME}
   \pi_{\mathrm{ME}}(\bm{s}) =&\frac{2^{\lfloor \frac{n}{2} \rfloor} + 2^{\lceil \frac{n}{2} \rceil} -1 }{2^n}  \nonumber \\
&+ \frac{1}{2^{2n}} \sum_{k \in \X^n} \sum_{l,m \in \X_*^{n}} 
 g(l,m)\, s_k s_{k\oplus l}^* s_{k\oplus l \oplus m} s_{k\oplus m}^*,
\end{align}
where \(g(l,m)\) is the coupling function~\eqref{gdef}.
From this expression, and proceeding  analogously to that in Sec.~\ref{sec3:fixed bipartition}, we obtain
\begin{equation} \label{meanME}
    \mu_{\mathrm{ME}}=\langle \pi_{\mathrm{ME}} \rangle 
    =\frac{2^{\lfloor \frac{n}{2} \rfloor} + 2^{\lceil \frac{n}{2} \rceil} -1 }{2^n},
\end{equation}
and
\begin{equation}
    \pi_{\mathrm{ME}}(\bm{s}) = \mu_{\mathrm{ME}} + \widetilde{\pi}_{\mathrm{ME}}(\bm{s}),
\end{equation}
where \(\widetilde{\pi}_{\mathrm{ME}}(\bm{s})\) denotes the second term in Eq.~\eqref{ExpressionPiME}.

By comparing~\eqref{mutypA} and~\eqref{meanME}, we get that
\begin{equation}
    \left( \mu_{\mathrm{ME}} \right)_{\mathrm{Hadamard}} < \left( \mu_{\mathrm{ME}} \right)_{\mathrm{Haar}},
\end{equation}
which confirms that Hadamard states exhibit, on average, a higher degree of entanglement than Haar-distributed states.

It follows from~\eqref{ExpressionPiME} that the variance of the average purity distribution is given by \(\sigma^2_{\mathrm{ME}} = \langle  \widetilde{\pi}_{\mathrm{ME}}^2 \rangle\). This quantity is explicitly evaluated in Appendix~\ref{Appendix_variances}, yielding 

\begin{equation}
    \sigma^2_{\mathrm{ME}} = \frac{c_q}{2^{3n}} f_{2*}(n), \label{sigmaME}
\end{equation}
where $c_q = (1+\delta_{q,2})$, for $q=2,3,\dots$, and
\begin{align} \label{def_f2*}
    f_{2*}(n) &= 4 \sum_{k,l\in\X_*^n} g(k,l)^2
    \nonumber\\
    &=\sum_{s,t=1}^{\lceil\frac{n}{2}\rceil}\! \binom{n}{s,t}^{\!\!\!-1} \!\left[ 
    \binom{\lfloor\frac{n}{2}\rfloor}{s} \binom{\lceil\frac{n}{2}\rceil}{t} 
    + \binom{\lfloor\frac{n}{2}\rfloor}{t} \binom{\lceil\frac{n}{2}\rceil}{s} \right]^2 \!\!.
\end{align}

Since $f_{2*}(n) < f_2(n)$, one also gets
\begin{equation}
    \left( \sigma_{\mathrm{ME}} \right)_{\mathrm{Hadamard}} < \left( \sigma_{\mathrm{ME}} \right)_{\mathrm{Haar}},
\end{equation}
for all $n\geq 4$ and all $P_q$-states with $q\neq 2$. Interestingly, the inequality is reversed for hypergraph states, $q=2$.

\begin{figure}[t]
    \centering
    \includegraphics[width=0.48\textwidth]{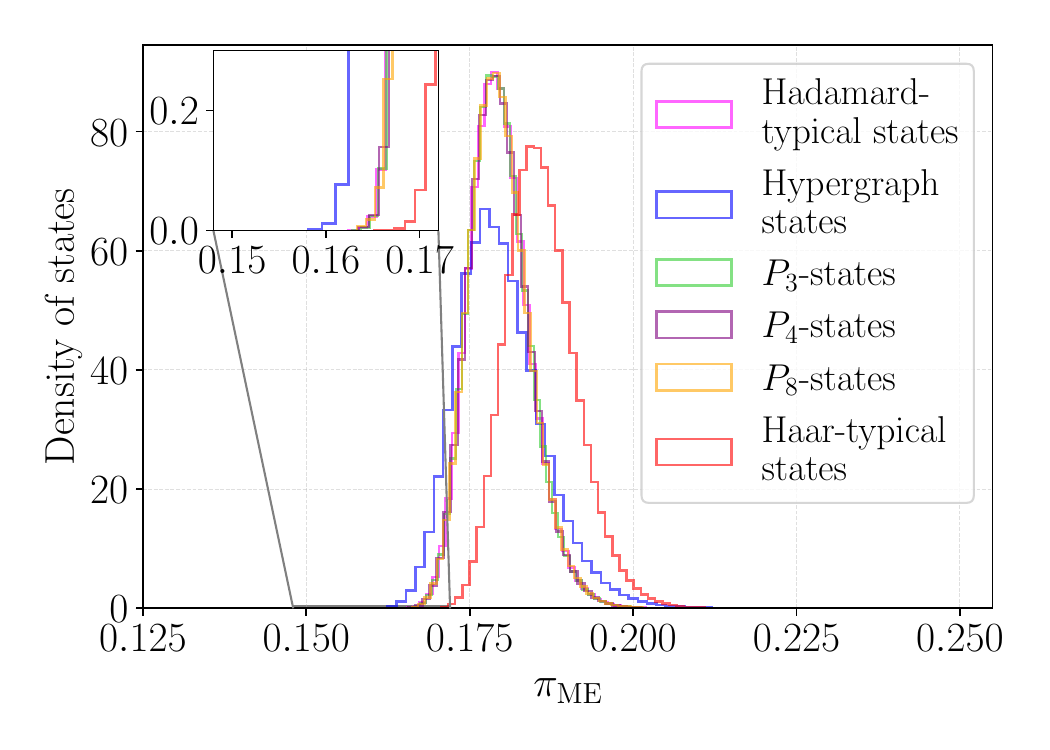}
    \caption{\justifying Histograms of the average purity density for $n=7$ qubits obtained from \(2.3 \times 10^6\) samples of different ensembles of states. The horizontal axis starts from $\pi_{\mathrm{ME}}=1/8=0.125$, which corresponds to the theoretical bound for the minimum, which would be saturated, if seven-qubit AME states existed. Using Eq.~\eqref{def_distance}, we observe that this value, is approximately 8.4 standard deviations from the mean value of the Hadamard-typical distribution.
    }\label{fig:comparisonMe}
\end{figure}

In Fig.~\ref{fig:comparisonMe}, we compare the distributions of the average purity across different classes of seven-qubit states obtained through numerical simulations (details in Appendix~\ref{Appendix_simulations}). The \(P_q\)-states exhibit statistical behavior similar to Hadamard-typical states up to the second cumulant. In contrast, hypergraph states ($q=2$) display significantly broader distributions, as predicted by
Eq.~\eqref{sigmaME}. All considered Hadamard states exhibit, on average, a higher degree of entanglement than Haar-typical pure states, consistently with Eq.~\eqref{meanME} and Eq.~\eqref{mutypA}. 

The broader distribution of hypergraph states increases the probability of sampling low average purity states, making them promising candidates in the search for MMES, as highlighted by the inset of Fig.~\ref{fig:comparisonMe}. Nevertheless, finding such states remains a formidable challenge, as indicated by the location of the theoretical bound for the minimum of the average purity (from which the horizontal axis in Fig.~\ref{fig:comparisonMe} starts). Specifically, we observe that, since
\begin{equation}
    k=\frac{\mu_{\mathrm{ME}}-2^{- \lfloor\frac{n}{2}\rfloor}}{\sigma_{\mathrm{ME}}}=\frac{(2^{\lfloor\frac{n}{2}\rfloor}-1)2^{\frac{n}{2}}}{\bigl(c_q f_{2*}(n)\bigr)^{\frac{1}{2}}},
    \label{def_distance}
\end{equation}
the theoretical minima, even in the case of hypergraph states, lie $k$ standard deviations below the average value, with
$k\simeq 2.9, 3.2, 7.5, 8.4$, and $19$ for $n=4,5,6,7$, and $8$, respectively.

\begin{figure}[t]
    \centering
    \includegraphics[width=0.46\textwidth]{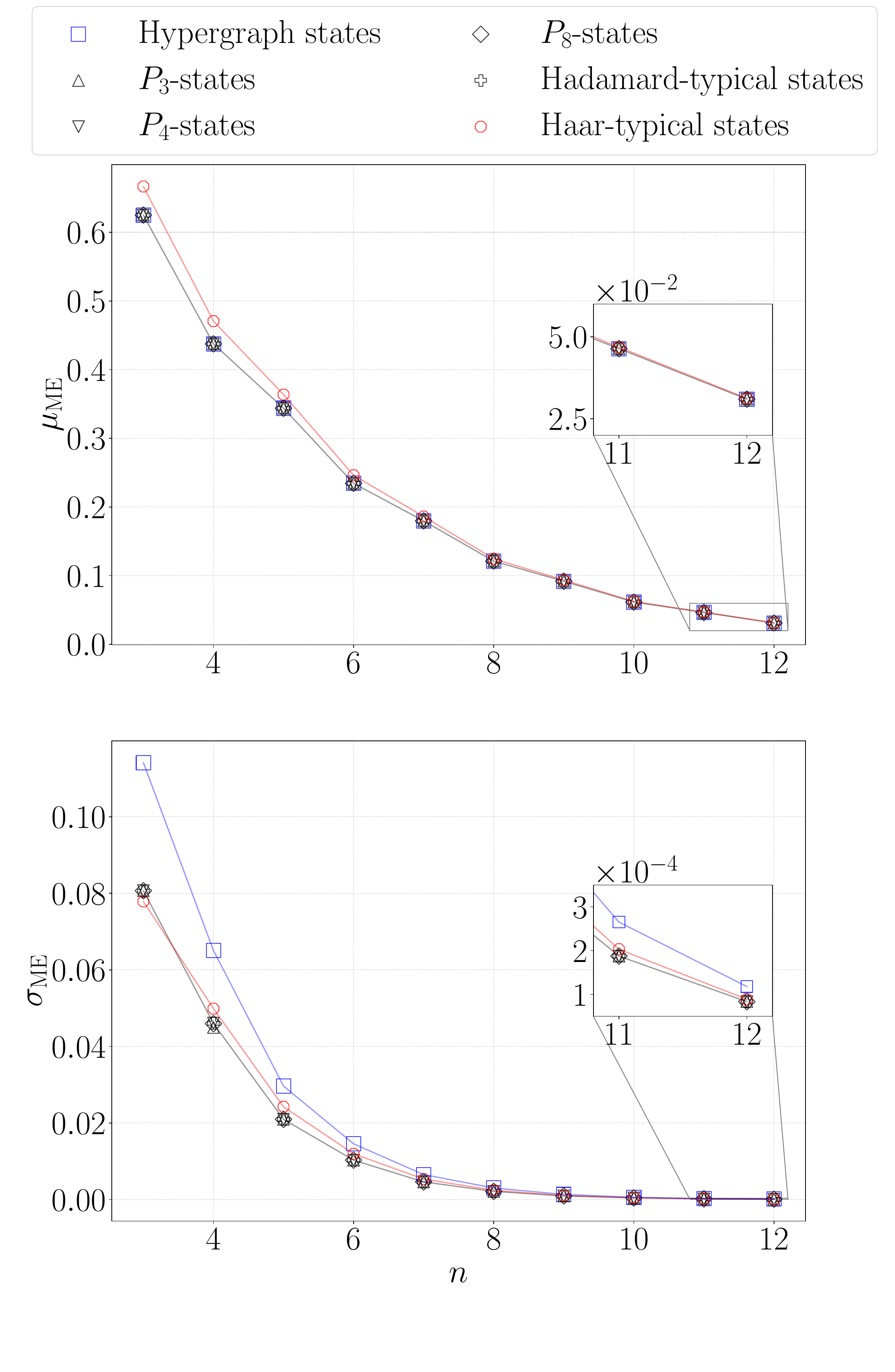}
    \caption{\justifying Mean and standard deviation of the average purity distribution as a function of system size, showing theoretical predictions (solid lines) and numerical results (scattered points) for Haar-typical states (red), hypergraph states (blue), Butson $P_3$-, $P_4$-, $P_8$-states and Hadamard-typical states (black).} \label{fig:predictions}
\end{figure}

We further extend our analysis to different system sizes by numerically simulating systems up to 12 qubits. In Fig.~\ref{fig:predictions}, we show analytical (solid lines) and numerical (scattered points) values of the mean and standard
deviation of the average purity distribution for all classes of states. Analytical results are given by: Eq.~\eqref{mutypA} and the square root of Eq.~\eqref{typicalSigmaME} for Haar-typical ensembles; Eq.~\eqref{meanME} and the square root of Eq.~\eqref{sigmaME} for all Hadamard ensembles. 
All the numerical results are in strong agreement with the analytical predictions, confirming the distinctive properties of \mbox{hypergraph states} ($P_2$-states).

\section{\label{sec:asymp}Large-System Limit}

As shown in Sec.~\ref{sec3:fixed bipartition}, the fluctuating part~\( \widetilde{\pi}_{\A}(\bm{s})\) 
of the purity of a fixed bipartition of a Hadamard state~\eqref{defPIa} consists of  \( O(2^{2n}) \) terms of order \( O(1/2^{2n}) \). 
For a large number $n$ of qubits, the purity distribution of Hadamard states referred to a fixed bipartition is close to a Gaussian distribution with mean given by Eq.~\eqref{meanA} and variance given by
Eq.~\eqref{varianceA}. 

The asymptotic mean of the purity distributions of Hadamard states~\eqref{meanA} coincides with that of Haar-typical states~\eqref{mutypA1},  
 \begin{equation}
        \mu_{\A} 
        \sim \frac{1}{2^{n_\A}} + \frac{1}{2^{n_{\bar{\A}}}}, 
    \end{equation}
as $n\to\infty$.
This occurs because the fluctuations in the moduli of the components a unit random vector become relatively small as the vector length increases, so a generic random vector effectively behaves like one with equal moduli.

Similarly, the asymptotic behavior of the variance~\eqref{varianceA} of the purity distribution for the Hadamard ensembles is
\begin{equation}
    \sigma_{\A}^2 \sim c_q \frac{2}{2^{2n}},
\end{equation}
as $n\to\infty$, where $c_q = 1+\delta_{q,2}$.
Thus, it is asymptotically equivalent to the variance~\eqref{sigmatypA} of the Haar ensemble, except in the special case $q=2$, corresponding to hypergraph states, where it is twice as large.

For large \(n\), the distribution of 
the average purity 
\(\pi_{\textrm{ME}}\) of Haar-typical states is also approximately Gaussian~\cite{Facchi10}. Its mean~\eqref{mutypA} behaves as
\begin{align}
        \mu_{\mathrm{ME}}  \sim \frac{1}{2^{\frac{n}{2}}}
        \begin{cases} 
            2 & \textrm{if } n \textrm{ even}\\
            \frac{3}{\sqrt{2}} &  \textrm{if } n \textrm{ odd}
        \end{cases}, 
        \label{eq:muMEmean}
    \end{align}
as $n\to\infty$.
Moreover, the function in~\eqref{def_f2}
has the following asymptotics~\cite{Facchi_classical}:
\begin{equation}
    f_2(n)
    \sim  3\sqrt{2}\left(\frac{3}{2}\right)^n
    = 3\sqrt{2}\,2^{\alpha n},
    \label{eq:f2as}
\end{equation}
as $n\to\infty$, where
\begin{equation}\label{def:alpha}
 \alpha = \log_2 \left(\frac{3}{2}\right) \simeq 0.58.
\end{equation}
Therefore, the variance $\sigma^2_{\mathrm{ME}}$ given by~\eqref{typicalSigmaME}, behaves as
\begin{align} 
\label{typicalSigmaMEas}
    \sigma^2_{\mathrm{ME}} 
    \sim 3\sqrt{2} \left(\frac{3}{16}\right)^n = 3\sqrt{2} \frac{1}{2^{(3-\alpha)n}}    .
\end{align}

Analogous considerations apply to the average purity of ensembles of Hadamard states, expressed by Eq.~\eqref{ExpressionPiME}, which also consists of at most \( 2^{2n} \) terms of order \( O(1/2^{2n}) \), and for large $n$ is close to a Gaussian 
with mean being given by Eq.~\eqref{meanME}, and  variance being provided by 
Eq.~\eqref{sigmaME} for all 
the ensembles of Hadamard $P_q$-states .

For large \( n \), the mean~\eqref{meanME} of the Hadamard ensembles has the same asymptotics~\eqref{typicalMeanME} of the Haar ensemble. Moreover, 
as derived in Appendix~\ref{Appendix_h},
the function $f_{2*}$ has the same asymptotic behavior~\eqref{def_f2} of $f_2$, namely,
\begin{align} 
    f_{2*}(n) 
    \sim  3\sqrt{2}\left(\frac{3}{2}\right)^n
    = 3\sqrt{2}\,2^{\alpha n},
\end{align}
as $n\to\infty$.
Therefore, in this limit,
the variance of the average purity distribution becomes
\begin{equation}\label{eq:sigmaME_approx_unif}
	\sigma^2_{\mathrm{ME}} 
    \sim 3\sqrt{2}\, c_q \left(\frac{3}{16}\right)^n = 3\sqrt{2}\, c_q \frac{1}{2^{(3-\alpha)n}}    ,
\end{equation}
with $c_q = (1+\delta_{q,2})$,
for all 
ensembles of Hadamard states in Table~\ref{tab:table1}.

Thus, in the large-system limit, all Hadamard states, except the hypergraph states, exhibit the same asymptotic purity distribution as Haar states. 

However, notice that Hadamard
states retain an entanglement advantage: the distance between the averages scales as
\begin{equation}
 \Delta\mu_{\mathrm{ME}} =(\mu_\mathrm{ME})_{\mathrm{Haar}}-(\mu_\mathrm{ME})_{\mathrm{Hadamard}} \sim \frac{1}{2^n},   
\end{equation}
while the variances scale as in Eqs.~\eqref{typicalSigmaMEas} and~\eqref{eq:sigmaME_approx_unif}, that is $\sigma_{\mathrm{ME}} = O((3/16)^{n/2})$. Thus, 
\begin{equation}
 \frac{\sigma_{\mathrm{ME}}}{\Delta\mu_{\mathrm{ME}}} = O\biggl(\Bigl(\frac{3}{8}\Bigr)^{\frac{n}{2}} \biggr) 
\end{equation}
 and the distributions of Hadamard states and Haar-typical states remain statistically distinct even for large systems. Furthermore, the hypergraph states retain a particular feature in the high-entanglement regime, showing increased probabilities even in the large-system limit. This distinctive scaling behavior, also highlighted by the inset of Fig.~\ref{fig:predictions}, makes hypergraph states uniquely valuable for exploring extreme entanglement even in large qubit systems.

\section{\label{Conclusion}Concluding remarks and outlook}
We systematically investigated the statistical properties of the purity and average purity  for various ensembles of Hadamard states of qubits. By extending the approach pioneered in Refs.~\cite{Facchi07,Facchi10,Facchi_classical}, we derived exact analytical expressions for the first and second cumulants of these distributions. We also performed extensive numerical simulations to validate these analytical results.
We showed that all considered Hadamard states exhibit on average a higher degree of entanglement than typical (Haar-distributed) states. 

Our analysis also revealed the unique properties of hypergraph states, whose purity fluctuations are significantly larger than those of the other ensembles of Hadamard $P_q$-states. The increased variance in the purity distribution of the hypergraph states suggests a higher probability of sampling states that approach the theoretical minimum of the average purity, even for large systems. This makes the hypergraph states promising candidates in the ongoing search for MMES.

An interesting future direction would be the extension of our cumulant analysis to higher-order moments, which could deepen the statistical characterization of purity distribution and potentially unveil additional features of Hadamard states compared to typical ones. Moreover, while MMES have been extensively studied among typical states~\cite{Facchi_frustration}, an extensive numerical and analytical analysis of MMES among Hadamard states, and particularly among hypergraph states, is still missing.

The identification and characterization of highly entangled multipartite states have implications beyond fundamental physics. Such states are crucial resources for quantum algorithms, particularly in quantum optimization protocols and certain quantum machine learning schemes where entanglement enables the efficient exploration of large solution spaces~\cite{PhysRevA.83.052313, 10.3389/fams.2021.716044, PhysRevA.111.022434, PhysRevLett.127.040501, PRXQuantum.6.010320, PhysRevLett.114.110504, ELAYACHI2025130666}. Additionally, understanding multipartite entanglement is relevant for developing and analyzing quantum error correcting codes ~\cite{Scott04, Raissi_2018, PhysRevA.101.042305}.

In summary, this work deepens our understanding of multipartite entanglement through statistical lens, revealing distinctive properties of Hadamard quantum states that suggest promising directions for future research.

\section{Acknowledgments}
We thank Chiara Macchiavello and Flavio Baccari for useful discussions and feedback on our work.

We acknowledge support from INFN through the project ``QUANTUM" and from the Italian funding within the ``Budget MUR - Dipartimenti di Eccellenza 2023–2027" - Quantum Sensing and Modelling for One-Health (QuaSiModO).
PF acknowledges support from the Italian National Group of Mathematical Physics (GNFM-INdAM) and from PNRR MUR project CN00000013-``Italian National Centre on HPC, Big Data and Quantum Computing``. GM acknowledges support from the University of Bari via the 2023-UNBACLE-
0244025 grant and from INFN through the project ``NPQCD". SP acknowledges support from PNRR MUR project PE0000023-NQSTI.  K{\.Z} acknowledges funding by the European Union under ERC
722 Advanced Grant TAtypic, Project No. 101142236. We acknowledge computational resources provided by the University of Bari and the INFN cluster
ReCaS~\cite{ReCaS}.

\appendix
\section{Computation of the variances\label{Appendix_variances}}
As shown in Sec.~\ref{sec3:fixed bipartition}, the variance of the purity distribution associated with Hadamard states and referred to a fixed bipartition is \(\sigma_A^2 = \left\langle \widetilde{\pi}_A^2 \right\rangle\),
where $\widetilde{\pi}_A$ is given in the second line of~\eqref{defPIa}.
Explicitly,
\begin{align}\label{eq:first_sigmaA}
    \sigma_{\A}^2 &= \frac{1}{2^{4n}} \biggl\langle \biggl( \sum_{k \in \X^n} \sum_{l \in \X_*^{\A}} \sum_{m \in \X_*^{\bar{\A}}} s_k s_{k \oplus l}^* s_{k \oplus l \oplus m} s_{k \oplus m}^* \biggr)  \nonumber \\
    &\quad  \times \biggl( \sum_{p \in \X^n} \sum_{v \in \X_*^{\A}} \sum_{r \in \X_*^{\bar{\A}}} s_p s_{p \oplus v}^* s_{p \oplus v \oplus r} s_{p \oplus r}^* \biggr) \biggr\rangle,
\end{align}
that is
\begin{align}
    \sigma^2_{\A} =& \frac{1}{2^{4n}} \sum_{k,p \in \X^n} \sum_{l,v \in \X^{A}_*} 
    \sum_{m,r \in \X^{\bar{\A}}_*} 
     \nonumber \\
    &\times 
    \bigl\langle  (s_k s^*_{k \oplus l} s_{k \oplus l \oplus m} s^*_{k \oplus m}) (s_p s^*_{p \oplus v} s_{p \oplus v \oplus r} s^*_{p \oplus r}) 
    \bigr\rangle.
    \label{eq:A2}
\end{align}

For the Hadamard ensembles considered here, the variables $s_k$ are
independent random phases, uniformly distributed on the regular polygon $P_q$, and $|s_k|^2=1$. Hence,
the expectation value of any product of such phases is non-zero only if, for
each index $k\in\X^n$, the number of occurrences of $s_k$ equals the number of
occurrences of $s_k^*$. Equivalently, the set of four indices carried by
the unconjugated phase factors must coincide with the set of four indices
carried by the conjugated phase factors.

In the present case, since $l,m$ (and $v,r$) are non-zero and have support on
complementary subsystems, no cancellation can occur within a single 4-phase factor.
Therefore, the only non-vanishing contributions arise when the indices of the
unconjugated phases in one factor are paired with the indices of the
conjugated phases in the other one, and vice versa. This can happen in
two ways, namely when $k=p\oplus v$ or when $k=p\oplus r$, which then fixes the remaining indices.

Let us illustrate this with the case $k=p\oplus v$. Then one unconjugated
phase factor from the first parenthesis in~\eqref{eq:A2} is paired with one conjugated phase
factor from the second parenthesis. For the remaining phase factors to give a
non-zero expectation value, their indices must also match, which imposes
$l\oplus m = v\oplus r$.
The remaining product is therefore
$s_{k \oplus l}^* s_{k \oplus m}^*
    s_{k \oplus v} s_{k \oplus v \oplus l \oplus m}$.
A non-vanishing contribution is obtained only when either \(l=v\) or \(m=v\).
However, since \(v\in \X_*^{\A}\), whereas \(m\in \X_*^{\bar{\A}}\), the latter
possibility is excluded. Thus the only admissible solution in this case is
\(l=v\), and consequently \(m=r\). The case \(k=p\oplus r\) is analogous.

Accordingly, the variance reads
\begin{align}
    \sigma^2_{\A} &= \frac{1}{2^{4n}} \!\sum_{k,p \in \X^n} \!\sum_{l,v \in \X^{A}_*}\! 
    \sum_{m,r \in \X^{\bar{\A}}_*}  (\delta_{k,p\oplus v}+\delta_{k,p\oplus r}) \delta_{l,v}\delta_{m,r}  \nonumber\\
    &= \frac{1}{2^{4n}} \!\sum_{p \in \X^n} \!\sum_{v \in \X^{A}_*} \!
    \sum_{r \in \X^{\bar{\A}}_*}  2 = 2 \frac{(2^{n_\A} - 1)(2^{n_{\bar{\A}}} - 1)}{2^{3n}}.
    \label{eq:expression_sigma}
\end{align}

For hypergraph states ($q=2$), whose coefficients are real, one has \(s_k^2 = s_k^{*2} = 1\). As a result,  additional non-zero contributions arise from direct pairings of the phases $s$  across the two parentheses. This happens when  $k = p$ or $k = p \oplus v \oplus r$ which, similarly to the previously discussed cases, fix all other indices. Together with the conditions originating from the pairing between phase factors $s$ and  conjugated phase factors $s^*$, these yield four distinct matching configurations, leading to the contribution 
\begin{equation}
    (\delta_{k,p\oplus v}+\delta_{k,p\oplus r}+\delta_{k,p}+\delta_{k,p\oplus v\oplus r}) \delta_{l,v}\delta_{m, r}.
\end{equation}
This effectively doubles the number of contributing configurations with respect to the other Hadamard ensembles.

As a result, the variance of the purity distribution for Hadamard $P_q$-states and referred to a fixed bipartition is  given  by 
\begin{equation}
    \sigma_{\A}^2 = 2 c_q \frac{(2^{n_\A} - 1)(2^{n_{\bar{\A}}} - 1)}{2^{3n}},
\end{equation}
with 
\begin{equation}
    c_q = (1+\delta_{q,2}),
\end{equation} 
for all $q\geq 2$, including Hadamard-typical states ($q=\infty$).

A similar analysis applies to the computation of the variance of the average purity distribution. As showed in Sec.~\ref{sec4:MEcumulants}, this quantity is given by \(\sigma^2_{\mathrm{ME}} = \left\langle \left( \widetilde{\pi}_{\mathrm{ME}} \right)^2 \right\rangle\), which explicitly reads
\begin{align}
    \sigma^2_{\mathrm{ME}} =& \frac{1}{2^{4n}} \sum_{k,p \in \X^n} \sum_{l,m,v,r \in \X^n_*} g(l,m)\, g(q,r)
     \nonumber \\
    &\times 
    \bigl\langle  s_k s^*_{k \oplus l} s_{k \oplus l \oplus m} s^*_{k \oplus m}\, s_p s^*_{p \oplus v} s_{p \oplus q \oplus r} s^*_{p \oplus r} 
    \bigr\rangle.
\end{align}

As in the fixed bipartition case, the non-vanishing contributions to the variance arise from matching index configurations between the \(s\)-terms and their conjugates. However, in this case, the conditions \(k = p \oplus v\) or \(k = p \oplus r\) do not fully fix the remaining indices. For instance, if \(k = p \oplus v\), then the relevant term becomes \(s_{k \oplus l}^* s_{k \oplus m}^* s_{k \oplus v} s_{k \oplus v \oplus l \oplus m}\), which contributes non-trivially when either \(l = v\) or \(m = v\). Since \(l, m, v, r \in \X^n_*\), both configurations are admissible, resulting in four valid matchings for all $P_q$-states with $q\geq3$.
In contrast, for $q=2$, non-vanishing contributions also arise from configurations in which the \(s\)-terms of both factors are directly matched. This leads to eight distinct non-vanishing terms, thereby doubling the number of contributing terms compared to the other ensembles.

Summarizing the above, we find that
\begin{equation}
 \sigma^2_{\mathrm{ME}} =   \frac{4 c_q}{2^{3n}}  \sum_{l,m \in \X_*^{n}}
    g(l,m)^2,
    \label{eq:sigma3MEder}
\end{equation}
with  $c_q = (1+\delta_{q,2})$.
We get
\begin{align} 
    f_{2*}(n) &= 4 \sum_{k,l\in\X_*^n} g(k,l)^2
    = 4 \sum_{k,l\in\X_*^n} \delta_{k\wedge l, 0}\, \hat{g}(|k|,|l|)^2
    \nonumber\\
    &= \sum_{1\leq s,t\leq n}4 \hat{g}(s,t)^2\sum_{k,l\in\X_*^n} \delta_{k\wedge l, 0}\, \delta_{|k|,s} \delta_{|l|,t} 
    \nonumber\\
    &= \sum_{1\leq s,t\leq n}4 \hat{g}(s,t)^2\binom{n}{s,t} 
    \nonumber\\
    &=\sum_{s,t=1}^{\lceil\frac{n}{2}\rceil}\! \binom{n}{s,t}^{\!\!\!-1} \!\left[ 
    \binom{\lfloor\frac{n}{2}\rfloor}{s} \binom{\lceil\frac{n}{2}\rceil}{t} 
    + \binom{\lfloor\frac{n}{2}\rfloor}{t} \binom{\lceil\frac{n}{2}\rceil}{s} \right]^2 \!\!,
    \label{eq:f2*der}
\end{align}
where we used the straightforward identity 
\begin{align}
        \hat{g}(s,t) &= \frac{1}{2} \binom{n}{\lfloor \frac{n}{2} \rfloor}^{\!\!\!-1} \! \left[ \binom{n - s - t}{\lfloor \frac{n}{2} \rfloor - s} + \binom{n - s - t}{\lfloor \frac{n}{2} \rfloor - t} \right]
        \nonumber\\
        &= \frac{1}{2} \binom{n}{s,t}^{\!\!\!-1} \!\left[ 
    \binom{\lfloor\frac{n}{2}\rfloor}{s} \binom{\lceil\frac{n}{2}\rceil}{t} 
    + \binom{\lfloor\frac{n}{2}\rfloor}{t} \binom{\lceil\frac{n}{2}\rceil}{s} \right]
    \end{align}
 on the  function $\hat{g}(s,t)$ defined in~\eqref{gbardef}.
By plugging~\eqref{eq:f2*der} into~\eqref{eq:sigma3MEder} we obtain the result~\eqref{sigmaME}--\eqref{def_f2*} of the main text.

\section{Numerical simulations\label{Appendix_simulations}}
A Haar-typical pure quantum state of $n$-qubits can be obtained by acting with a random unitary operator, sampled according to the Haar measure, on a generic reference state. In other words, a typical state corresponds to a single row (or column) of a unitary matrix sampled according to the Haar measure~\cite{Zyczkowski01}. Based on this observation, we simulated typical states by generating full unitary matrices sampled from the Haar distribution using the algorithm presented in~\cite{Mezzadri07}. 

Hadamard states, in contrast, are characterized by a reduced number of parameters compared to Haar states, which makes their numerical simulation significantly simpler. 

To simulate a Hadamard pure state of an $n$-qubits system, we constructed an ordered list of $N=2^n$ phases ($s_k=e^{i\phi_k}$), uniformly sampled from the regular polygon defining the chosen class of Hadamard states, as reported in Table~\ref{tab:table1}. The vector of Fourier coefficient $(z_1,\dots,z_N)$ was then obtained using Eq.~\eqref{defUnif}. This procedure is computationally less demanding than generating a random unitary matrix. We performed simulations for \mbox{$P_2, P_3,P_4,P_8$-states} and Hadamard-typical states. 

For all mentioned ensembles of Hadamard states, as well ad for the Haar ensemble, we generated $2.3\cdot 10^6$ states for system sizes ranging from $n=3$ to $n=12$ qubits. In addition, due to the reduced dimension of the states space, we were able to generate all possible hypergraph states with $n=3,4$ and $P_3$- and \mbox{$P_4$-states} with $n=3$.

For both Hadamard and Haar states, we first computed the purity referred to the bipartition \mbox{$\A=\{1,...,\lfloor n/2\rfloor\}$}. Then, we generated the list of all the possible balanced bipartitions, computed the purity for each case and evaluated the average purity as their arithmetic mean.
Although the simulations of the purity referred to a fixed bipartition are specific to $\A=\{1,...,\lfloor n/2\rfloor\}$, the results are independent of this choice as the purity distributions are statistically independent of the specific bipartition~\cite{Facchi06}.

All numerical simulations were performed on the RECAS computing cluster \cite{ReCaS}, utilizing up to 64 cores and 32 GB of memory per job.

\section{Asymptotic behavior of $f_{2*}(n)$}
\label{Appendix_h}

The summation in the expression~\eqref{def_f2*} of $f_{2*}$ can be split as
\begin{equation} \label{eq:termSigmaME}
    4\Bigl(\sum_{l,m \in \X^{n}}
    g(l,m)^2 - 2\sum_{l \in \X^{n}} g(l,0)^2 + 1\Bigr),
\end{equation}
where we have used the fact that \(g(0,0) = 1\).
Therefore, we have
\begin{equation}
    f_{2*}(n) = f_2(n) - 2 h_2 (n) + 4,
\end{equation}
where \(f_2(n)\) is defined in Eq.~\eqref{def_f2} and
\begin{equation}
    h_2(n) = 4 \sum_{l \in \X^{n}} g(l,0)^2.
\end{equation}
This is the contribution with $t=0$ of the sum in the second line of~\eqref{def_f2}, namely,
\begin{equation} \label{def:h2}
    h_2(n) = \sum_{s=0}^{\lceil \frac{n}{2}\rceil} \binom{n}{s}^{-1} \left[ \binom{\lfloor \frac{n}{2}\rfloor}{s} + \binom{\lceil \frac{n}{2}\rceil}{s} \right]^2.
\end{equation}

\begin{figure}
\centering\includegraphics[width=0.48\textwidth]{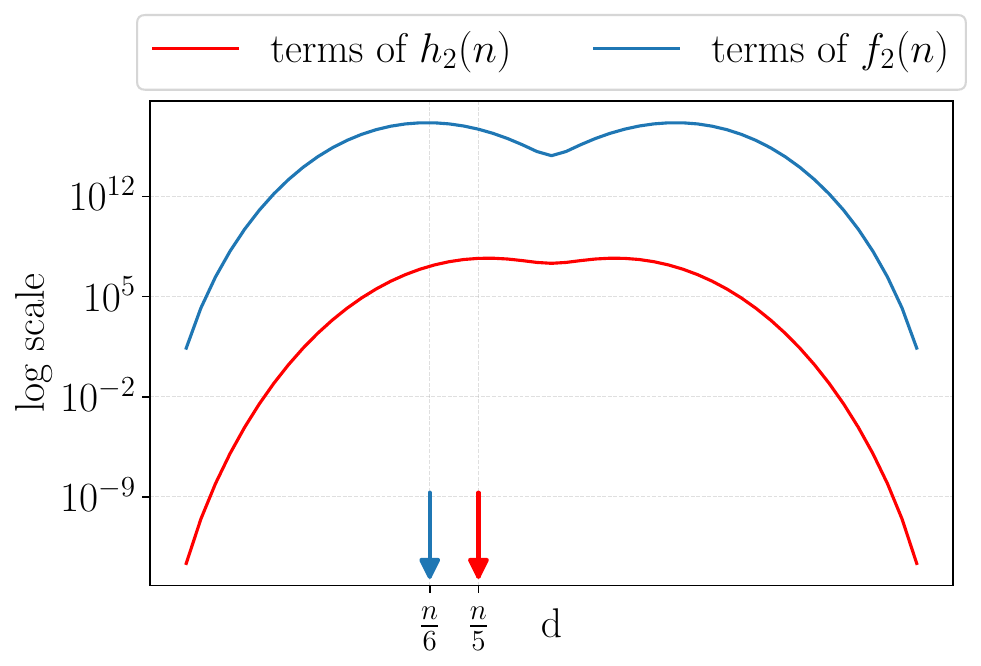}
	\caption{\justifying Comparison between the terms appearing in the sums defining  \( f_2(n) \) in~\eqref{eq:f2vsd} and \( h_2(n) \) in~\eqref{eq:h2vsd} versus the distance $d$~\eqref{eq:dist} between pairs of bipartitions, in logarithmic scale. We set $n=100$. The maxima of the terms of both sums, expressed as fractions of \( n \), are highlighted.}
	\label{fig:comp_f_h}
\end{figure}

We derive the asymptotic behavior of the function $h_2(n)$ 
defined in Eq.~\eqref{def:h2}.
Let us consider the case of even \( n \): 
\begin{equation}
    h_2(n) = 4 \sum_{0 \leq s \leq \frac{n}{2}} \binom{n}{s}^{-1} \binom{\frac{n}{2}}{s}^2.
\end{equation} 
This summation  depends on the ratio $s/n$, which we can define as a new variable
\begin{equation}
    x=\frac{s}{n}.
\end{equation}
By Stirling approximation $n!\sim(n/e)^n\sqrt{2\pi n}$ we obtain
\begin{align}
    h_2(n)&\sim \frac{4}{\sqrt{2\pi n}}\sum_{0 \leq s \leq \frac{n}{2}}  \sqrt{\frac{1-x}{x} }\frac{1}{1-2x} \, e^{n\left[H(2x)-H(x)\right]}\nonumber \\
    &\sim4{\sqrt{\frac{n}{2\pi} }}\int_{0}^{\frac{1}{2}}\mathrm{d}x\,  \sqrt{\frac{1-x}{x} }\frac{1}{1-2x} \, e^{n\left[H(2x)-H(x)\right]},
\end{align}
where 
\begin{equation}
    H(x)=-x\ln x-(1-x)\ln(1-x)
\end{equation}
is the Shannon entropy of $x$.
It is easy to see that the exponent $H(2x)-H(x)$ takes its maximum at $x_0=(2-\sqrt{2})/4$, and using the saddle point approximation we get
\begin{align}
    h_2(n)&\sim 4\sqrt{\frac{n}{2\pi }}
    \sqrt{\frac{1-x_0}{x_0} }\frac{1}{1-2x_0}
    e^{n\left[H(2x_0)-H(x_0)\right]} \nonumber \\
    &\times \int_{0}^{1/2}\mathrm{d}x\,e^{\frac{n}{2}\left[4H''(2x_0)-H''(x_0)\right]\left(x-x_0\right)^2}\nonumber\\
    &\sim 4\sqrt{\frac{n}{2\pi }}
    (2+\sqrt{2})    
    e^{n\log \left(\frac{1+\sqrt{2}}{2}\right)}\int_{\mathbb{R}}\mathrm{d}x\,e^{-4\sqrt{2} n x^2}
    \nonumber \\
    &= 2^{1/4} (2+\sqrt{2}) \left(\frac{1+\sqrt{2}}{2}\right)^n 
    \nonumber \\
    &= 2^{1/4} (2+\sqrt{2})\,2^{\gamma n},
    \label{eq:saddlepoint}
\end{align}
where
\begin{equation}
    \gamma =  \log_2 \left(\frac{1+\sqrt{2}}{2}\right) \simeq 0.27.
\end{equation}
For odd $n$, the asymptotic expression~\eqref{eq:saddlepoint} acquires the additional factor $(4+3\sqrt{2})/8$.

Therefore, the function $h_2(n)$ is asymptotically negligible with respect to the asymptotic behavior~\eqref{eq:f2as} of $f_2(n)$, since $\alpha > 2\gamma$.
It follows that, for large $n$, $f_{2*}(n)$ is asymptotically equivalent to $f_2(n)$, namely,
\begin{align} 
    f_{2*}(n) 
    \sim  3\sqrt{2}\left(\frac{3}{2}\right)^n
    = 3\sqrt{2}\,2^{\alpha n},
\end{align}
as $n\to\infty$.\\

In Ref.~\cite{Facchi_classical}, it was shown that the sum~\eqref{def_f2} defining $f_2(n)$ can be expressed in terms of the distance
\begin{equation}
    d= |A\cap \bar{B}|
    \label{eq:dist}
\end{equation}
between two bipartitions $(A,\bar{A})$ and $(B,\bar{B})$, as
\begin{align} 
    f_2(n) =& \binom{n}{\lfloor \frac{n}{2}\rfloor}^{\!\!\!-1} \sum_{d = 0}^{\lfloor\frac{n}{2}\rfloor} \binom{\lfloor \frac{n}{2}\rfloor}{d} \binom{\lceil \frac{n}{2}\rceil}{d} 2^{\frac{n}{2} +1 }  \nonumber \\
    &\qquad\qquad\qquad \times \left[ 2^{\frac{n}{2} - 2 d} + 2^{-(\frac{n}{2} - 2d)} \right].
    \label{eq:f2vsd}
\end{align}

The corresponding expression for $h_2(n)$ reads
\begin{align} 
    h_2(n) =&  \binom{n}{\lfloor \frac{n}{2}\rfloor}^{\!\!\!-1} \sum_{d = 0}^{\lfloor\frac{n}{2}\rfloor} \binom{\lfloor \frac{n}{2}\rfloor}{d} \binom{\lceil \frac{n}{2}\rceil}{d}   \nonumber \\
    &\qquad\quad\quad \times \left[ 
    (2^{\lfloor \frac{n}{2}\rfloor} + 2^{\lceil \frac{n}{2}\rceil} )
    2^{- d} + 2^{d+1} \right].
    \label{eq:h2vsd}
\end{align}
The comparison between the terms in the sums~\eqref{eq:f2vsd} and~\eqref{eq:h2vsd} is shown in Fig.~\ref{fig:comp_f_h} in the deep asymptotic regime, at $n=100$, confirming the asymptotic dominance of $f_2$ over $h_2$.
\vspace{\fill}

\bibliography{biblio}
\end{document}